\documentclass[sigconf]{acmart}

\makeatletter                   
\def\mdseries@tt{m}             
\makeatother                    

\usepackage[plain]{fancyref}
\usepackage[draft=true]{minted} 
\usepackage{color}
\usepackage{booktabs}
\usepackage{flushend}
\hypersetup{
    colorlinks=true,
    linkcolor=blue,
    filecolor=red,      
    urlcolor=magenta,
    breaklinks=true,            
}
\usepackage{breakurl}           
\usepackage[utf8]{inputenc}

\begin{document}
\sloppy                         

\title{How is Performance Addressed in DevOps?\\ A Survey on Industrial Practices}
\renewcommand{\shorttitle}{How is Performance Addressed in DevOps?}

\author{Cor-Paul Bezemer$^1$, Simon Eismann$^2$, Vincenzo Ferme$^3$, Johannes Grohmann$^2$, Robert Heinrich$^4$, Pooyan Jamshidi$^5$, Weiyi Shang$^6$, André van Hoorn$^7$, Monica Villaviencio$^8$, Jürgen Walter$^2$, Felix Willnecker$^9$}
\affiliation{$^1$University of Alberta (Canada) $^2$University of Würzburg (Germany) $^3$Software Institute, USI-Lugano (Switzerland)\\
$^4$Karlsruhe Institute of Technology (Germany) $^5$University of South Carolina (USA) $^6$Concordia University (Canada)\\
$^7$University of Stuttgart (Germany) $^8$ESPOL (Ecuador) $^9$fortiss GmbH (Germany)\\
rgdevops@spec.org}

\renewcommand{\shortauthors}{CP. Bezemer et al.}
\date{}

\hyphenation{Dev-Ops}

\acmConference[Preprint]{}{}

\begin{abstract}
DevOps is a modern software engineering paradigm that is gaining widespread adoption in industry. The goal of DevOps is to bring software changes into production with a high frequency and fast feedback cycles. This conflicts with software quality assurance activities, particularly with respect to performance. For instance, performance evaluation activities\,---\,such as load testing\,---\,require a considerable amount of time to get statistically significant results. %

We conducted an industrial survey to get insights into how performance is addressed in industrial DevOps settings. In particular, we were interested in the frequency of executing performance evaluations, the tools being used, the granularity of the obtained performance data, and the use of model-based techniques. The survey responses, which come from a wide variety of participants from different industry sectors, indicate that the complexity of performance engineering approaches and tools is a barrier for wide-spread adoption of performance analysis in DevOps. The implication of our results is that performance analysis tools need to have a short learning curve, and should be easy to integrate into the DevOps pipeline.

\end{abstract}

\maketitle

\pagebreak

\section{Introduction}
\label{sec:intro}

\begin{table*}[t]
\centering
\caption{A summary of the implications of our findings}
\label{tab:implications}
\resizebox{0.99\textwidth}{!}{
\begin{tabular}{ll}
\toprule
\textbf{1. The complexity of performance engineering approaches is a barrier for wide-spread adoption by practitioners.} & Section \ref{sec:impl_complexity} \\

\textbf{2. Performance engineering approaches must be lightweight.} & Section \ref{sec:impl_lightweight} \\
\textbf{3. Performance engineering approaches must smoothly integrate with existing tools in the DevOps pipeline.} & Section \ref{sec:impl_integrate} \\
\bottomrule
\end{tabular}
}
\end{table*}

DevOps is a modern software engineering paradigm that aims to reduce the time between changing software and delivering these changes into production with high quality~\cite{zhu2016devops}. This reduction in delivery time is achieved through organizational changes that bring together development and operations teams and processes with a high degree of automation, e.g., via continuous delivery (CD) pipelines and quality gates~\cite{Humble2010CD}.  

One of the most important quality aspects of a software system is performance.
The performance of a system can be described as several system properties that concern the system's timeliness and use of resources~\cite{Jain1991TheArtOfComputerSystemsPerformanceAnalysis}.  Common performance metrics are response time, throughput, and resource utilization. Performance requirements for software systems are typically defined by setting upper and/or lower bounds for these and other metrics. In order to ensure that such performance requirements can be met, several activities are required during the development and operation of these systems~\cite{Bondi2014PeformanceBook}. A common distinction is made between model-based activities, such as prediction using performance models~\cite{Cortellessa2011Book}, and measurement-based activities, such as load testing~\cite{Jiang2015LoadTestingSurvey} and monitoring~\cite{Heger:2017:APM:3030207.3053674}. Historically, performance-related activities in software development and operations were tackled independently from each other, but the newly emerging DevOps concepts require and enable a tighter integration between both activity streams~\cite{SPEC-RG-2015-001-DevOpsPerformanceResearchAgenda}. 

In our prior work~\cite{SPEC-RG-2015-001-DevOpsPerformanceResearchAgenda}, we discussed how existing solutions could support this integration, as well as open research challenges in the area of performance evaluation in DevOps. Despite the widespread adoption of DevOps practices and technologies, there are still many unanswered questions about DevOps. 

In this paper, we discuss our survey on the current state-of-practice of addressing performance concerns in industrial DevOps applications. 
Prior empirical studies show that the adoption of DevOps correlates with positive software quality outcomes~\cite{vasilescu2015quality}. Also, in the open source community, the usage of DevOps and continuous integration (CI) leads to more frequent releases~\cite{hilton2016usage}. However, these studies do not present the current practice of performance engineering in DevOps applications.
Our survey is the first to focus on performance engineering practices in a DevOps setting.

\pagebreak

In particular, we focus on the following aspects:

\begin{enumerate}
    \item How often are performance evaluations of applications developed using DevOps conducted in industry?
    \item Which performance evaluation tools are being used in the CD pipeline?
    \item What is the granularity of the analyzed performance data?
    \item Are performance models used in the CD pipeline?
\end{enumerate}

Our study reveals that automatic performance evaluations are usually not integrated into automatic delivery pipelines and not performed regularly. In addition, performance modeling is not applied in most companies. In this study, we observed that diagnosing performance issues is typically performed based on ``human intuition''~\cite{kaldor2017canopy}: engineers investigate hypotheses about what might have gone wrong in the system using data analytics to draw a conclusion about the observed performance issue. 


The remainder of this paper is structured as follows. Section~\ref{sec:related} provides an overview of related work, focusing on surveys about DevOps practices. 
Section~\ref{sec:methodology} presents details about our methodology, including the survey design. 
The main results of our survey are discussed in Section~\ref{sec:core_find}. Section~\ref{sec:implications}, discusses the main implications (which are summarized in Table~\ref{tab:implications}) of our study. In Section~\ref{sec:threats}, 
we discuss the threats to validity of our study. 
In Section~\ref{sec:concl}, we conclude the paper.

\section{Related Work}
\label{sec:related}
Others have performed prior surveys to assess the state-of-practice of DevOps in industry. Several of these surveys were conducted by corporations that sell DevOps solutions to companies.
While some surveys touched briefly upon the topic of software performance in DevOps, none of them focused on getting an in-depth overview of how performance engineering is applied in DevOps.
Prior surveys on the organizational impact of applying DevOps in industry~\cite{Puppet,Logz.io,Erich:2017:QSD:3129397.3129398}, assessed the DevOps adoption over different years~\cite{Puppet} and the types of tools and techniques used in DevOps pipelines~\cite{CATechnologies}. These surveys concluded from practitioner responses that DevOps has an increasingly large impact in industry. These prior surveys focused on the used tools, and underline how these tools are usable to optimize certain businesses and technology goals, such as improving software performance. 
In particular, software performance is discussed as one of the main drivers for using DevOps~\cite{CATechnologies,Atlassian2017}. Other drivers of the DevOps movement are: ``more efficient time-to-production for new software; a better collaboration between IT and other lines of business; and more consistent and higher quality software deployments''~\cite{KMSTechnology}. Overall, the surveys conclude that the DevOps trend is substantial and long-term. 
Puppet~\cite{Puppet} collected responses from 3200 surveyed practitioners, and reported that the percentage of teams that use DevOps (compared to other IT-related teams) increased from 16\,\% in 2014 to 27\,\% in 2017. As these percentages show, DevOps can still be considered relatively new and far from being applied widely in industry, as also reported by Logz.io~\cite{Logz.io} and Erich et al.~\cite{Erich:2017:QSD:3129397.3129398}. 
CA Technologies~\cite{CATechnologies} discusses the findings from an audience of 1425 senior IT and line-of-business executives and reports on the most critical DevOps demand drivers and tools, along with DevOps benefits and the factors that are driving DevOps. It is interesting to notice that improving the quality and performance of the applications is the top driver, with 42\,\% of the participants agreeing on this. Tool-wise, application performance management and monitoring (APM)~\cite{Heger:2017:APM:3030207.3053674} tools are perceived as the most important tools for DevOps by 38\,\% of the participants, while 37\,\% of the participants consider performance testing tools as critical. 
KMS Technology~\cite{KMSTechnology} surveyed 200 IT practitioners who were involved in transitioning to DevOps, and reported that 51\,\% had a very positive impression, and 79\,\% had achieved their desired goals. They also reported that the most significant challenge during the transition was the limited skill set and knowledge about DevOps among in-house IT staff (28\,\%). The second biggest challenge was a lack of support from the executive staff (23\,\%), followed by an inability to agree on and/or articulate the goals of the transition (18\,\%).

In addition, prior surveys of practitioners targeted the industrial adoption of performance testing~\cite{TechBeacon} and CI~\cite{hilton2016usage}. 
The report by TechBeacon~\cite{TechBeacon} is indirectly related to DevOps because the survey assessed performance engineering practices throughout the software development life cycle, and reported that 62\,\% of the participants agreed that performance engineering is important for DevOps. 
Hilton et al.~\cite{hilton2016usage} studied the barriers that developers face when using CI, and reported that the complexity of CI tools and flaky tests are important barriers for effective DevOps integration.

\section{Methodology}
\label{sec:methodology}
This section describes the design of the survey, the way in which it was advertised, 
and the profile of the participants.

\subsection{Survey Design}
The survey design follows the guidelines for conducting surveys in 
software engineering by Linåker et al.~\cite{Linaker2015}.
We designed our web survey to answer how industry addresses performance in DevOps processes.
 
Our survey contained 58 questions, divided into three parts: 1) questions about the participants' professional information (11 questions); 2) questions about development process models and team organization (30 questions); and 3) questions about performance assessment and evaluation (17 questions).

Based on the four aspects that are specified in Section~\ref{sec:intro}, we defined the target audience for the survey mainly as DevOps engineers, software architects, software developers, software operation engineers, software performance testers, and software consultants with a focus on performance engineering at software vendors and consultant companies worldwide.

We developed a set of initial hypotheses, such as on the frequency of performance evaluations, the applied tools and the acceptance of performance models.
Based on the set of hypotheses, we derived a questionnaire plan, consisting of survey goals, such as ``Measure capabilities of monitoring tools'' or ``Measure the completeness of the continuous delivery pipeline''. 
Each goal is composed of a set of concrete questions by which we want to answer the corresponding goal.
Additionally, the survey design aims not only at describing ``what'' happens, but also at answering ``why'' it happens in order to conduct an explanatory study as opposed to just being descriptive. 


In order to enable comparison, we aimed at minimizing free text questions and introduced single and multiple choice questions as well as Likert scales as often as possible to order the choices. Questions with ordered choices are less difficult to answer for participants and easier to analyze for researchers than unordered ones~\cite{burns2008basic,fink2012conduct,salant1994conduct}. 

\subsection{Survey Context and Advertisement}
\label{sec:advertisement}
We advertised the link of the survey through industry-related mailing lists
such as 
the SPEC (Standard Performance Evaluation Corporation)\footnote{\url{https://www.spec.org/}} mailing list,   
social media, related events such as DevOpsDays\footnote{\url{https://www.devopsdays.org/}} and links in online computer magazines and blogs. 
In addition, the request for participation in the survey was spread via the authors' network of industry contacts.
 
The data collection was conducted between May 2016 and March 2017. By the time this article was written, 26 full responses (all questions answered by participants) and 108 partial responses (a part of the questions answered) were gathered. 
The following sections of this paper are based on the 26 full responses only. \\ 

\subsection{Survey Participants}
The collected responses cover a wide range of education levels, processes, roles, work experiences and company sizes.

Approximately 85\,\% of the participants have a university degree (i.e., a Bachelor's degree (35\,\%), a Master's degree (25\,\%), or a Ph.D. (25\,\%)), while the other 15\,\% of the participants hold a high school degree.

There is a variety of job positions represented in the sample; however, more than a half of the participants describe themselves as software developers, and less than 10\,\% as DevOps engineer or performance engineer.

Most (56\,\%) of the participants have 1 to 3 years of working experience in their current position, while 22\,\% have 3 to 5 years of experience, and 22\,\% have 5 or more years of experience.
 
The participants work in companies that have between 100 and 999 employees (42\,\%), between 10 and 99 employees (31\,\%), and between 1,000 and 9,999 employees (19\,\%). The remaining participants work at companies that have less than 10 employees or more than 10,000 employees (8\,\%).

Most participants apply continuous integration 
(54\,\%)
while continuous deployment 
(12\,\%)
and continuous provisioning 
(4\,\%)
are applied less frequently. Continuous integration is often 
(38\,\%)
applied in combination with agile processes, such as Scrum. 
Most participants 
(54\,\%) 
use real-time data for process improvement.

\section{The Main Results of Our Survey}
\label{sec:core_find}
In this section, we present the main results of our survey. The complete questionnaire, raw response data, and a more detailed analysis are publicly available online~\cite{Survey}. 

\subsection{Performance evaluations are not regularly conducted in most companies}
\label{sec:regularly}
\begin{figure}
   \begin{minipage}[t]{\linewidth}
        \includegraphics[scale=0.37, trim= 120 250 0 120]{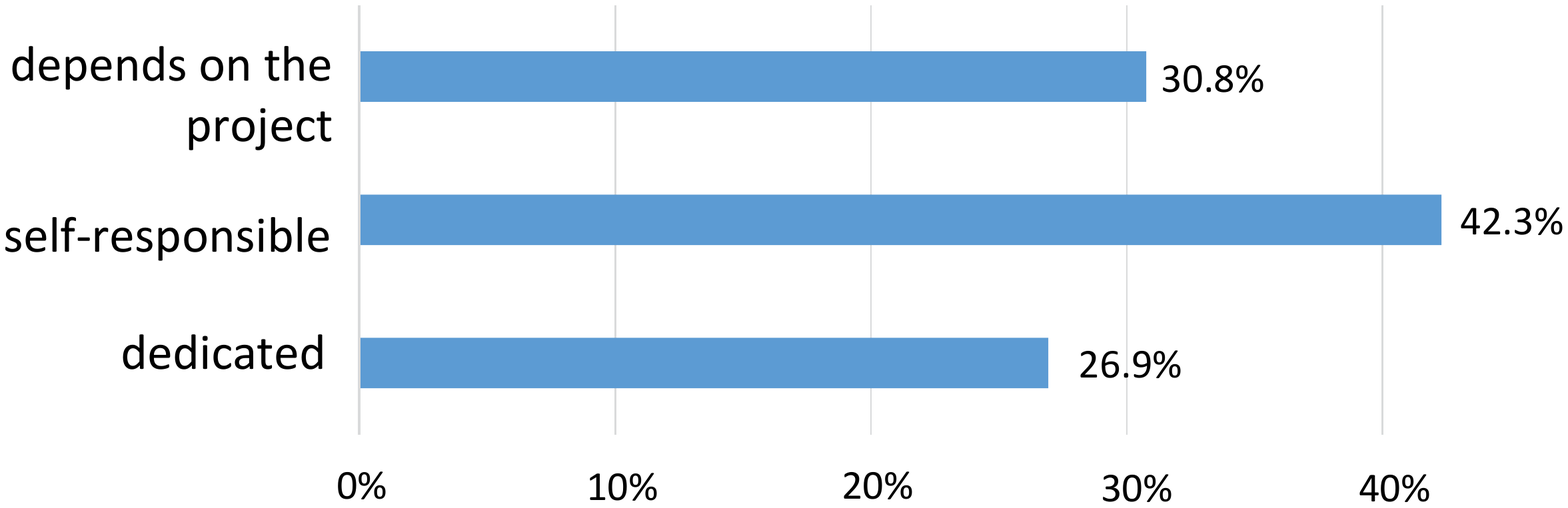}
        \captionof{figure}{Responsibility for performance evaluation} 
        \label{fig:responsabilities}
    \end{minipage}%
    \end{figure}

Approximately one third of the participants conducts performance evaluations on a regular basis (19\,\% continuously, 8\,\% daily, and 8\,\% weekly). The other participants conduct performance evaluations monthly (8\,\%), quarterly (27\,\%), yearly (12\,\%), less than yearly (8\,\%), or never (12\,\%). In addition, 50\,\% of the participants spend less than 5\,\% of their time, and only 20\,\% spend more than 20\,\% of their time on performance. 26\,\% of the participants report that performance evaluations are assigned to dedicated persons or teams; 41\,\% report to be in charge themselves (see Fig.~\ref{fig:responsabilities}).

\subsection{Jenkins is by far the most widespread CI solution}
\label{subsubsec:jenkins}
\begin{figure}
    \begin{minipage}[t]{\linewidth}
        \includegraphics[ scale=0.37, trim= 120 130 0 140]{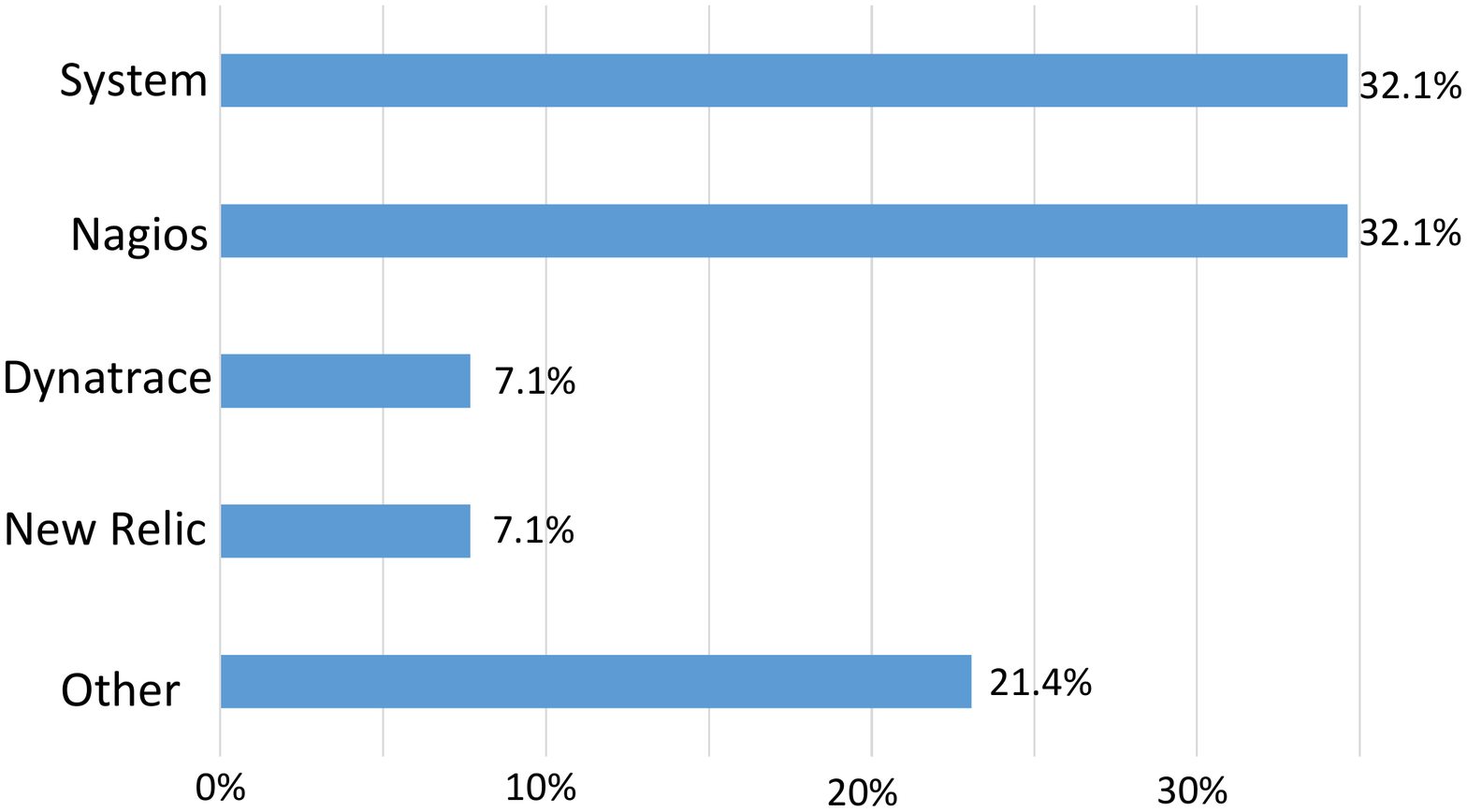}          
        \captionof{figure}{Employed performance evaluation tools}           
        \label{fig:tools}
    \end{minipage}
    \end{figure}

There exists a wide variety of tools that support the continuous integration pipeline. Not surprisingly, version control systems (VCSs) are used by all surveyed practitioners. The vast majority uses Git 
(77\,\%)
and/or SVN 
(38\,\%)
as VCS. Jenkins is the most popular ``end-to-end'' solution for CI. 
A majority of 77\,\% of the practitioners use Jenkins for continuous builds and 
65\,\% of the practitioners use Jenkins to deploy their software. 
Surprisingly,
50\,\% of the practitioners use SSH as a deployment system, beating Puppet (31\,\%) at the third place. The relatively heavy use of SSH suggests that CI solutions such as Jenkins cannot yet fulfill all wishes of practitioners, e.g., because such solutions are not capable of working with legacy code. To monitor performance, practitioners tend to rely on lower level system tools (35\,\%), such as top, or Nagios (35\,\%). APM tools (which are advertised as tools that support CI) are hardly used by practitioners (see Fig.~\ref{fig:tools}).
   
       \begin{figure}
    \begin{minipage}[b]{\linewidth}
        \includegraphics[scale=0.35, trim= 50 140 0 100]{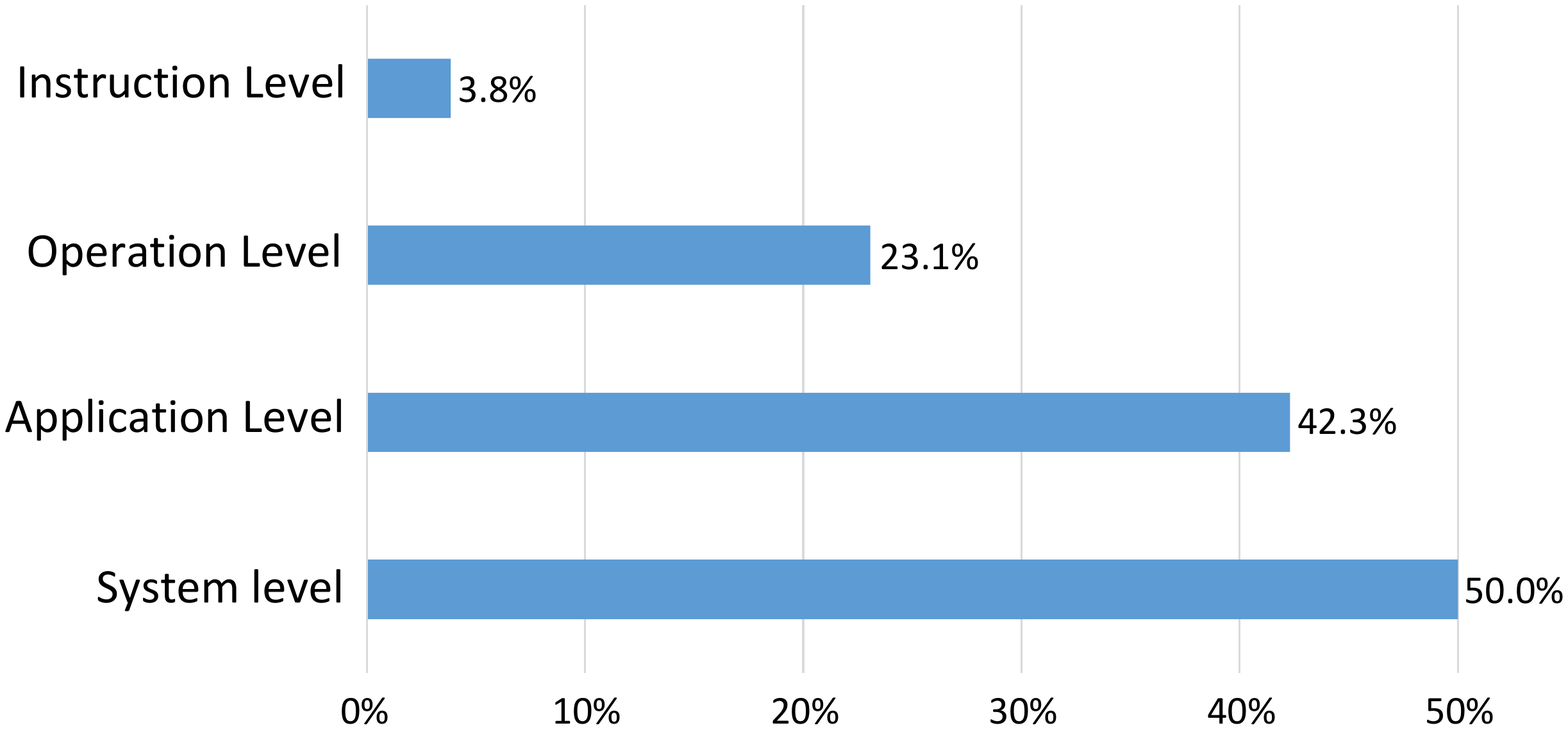}
        \captionof{figure}{Granularity of system monitoring}   
        \label{fig:monitoring_granularity}
    \end{minipage}
   \end{figure}
   
   \begin{figure}
   \begin{minipage}[b]{\linewidth}
        \includegraphics[
        scale=0.33, trim= 50 100 0 140
        ]{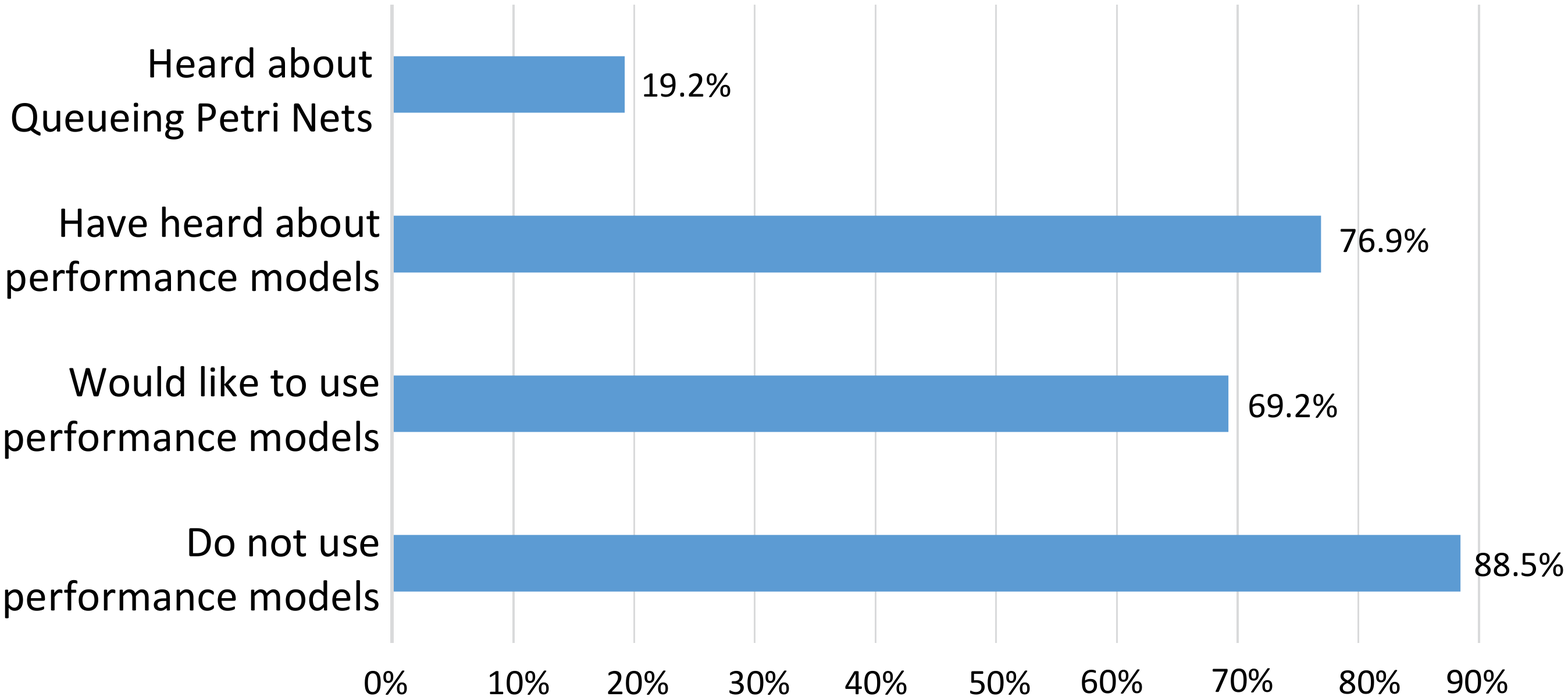}
        \captionof{figure}{Adoption of and attitude towards performance models}   
        \label{fig:perf_models}
    \end{minipage}
   \end{figure}
    
\subsection{Application-level monitoring is mostly done in an ad-hoc manner}
\label{subsubsec:apm}
Even though 70\,\% of the participants reported to have access to monitoring data, the responses on how their systems are monitored were surprising (see Fig.~\ref{fig:monitoring_granularity}). While monitoring system-level (and infrastructure) metrics is common, hardly any monitoring is conducted at higher levels, in particular, at the application-level (e.g., using application-internal metrics). The lack of application-level monitoring is reflected by both the reported granularity of measurements and the used tools. The granularity of monitoring is mentioned with decreasing occurrence from system level 
(50\,\%),
over application level 
(42\,\%)
and operation level 
(23\,\%),
to instruction level 
(4\,\%). 
Typical system-level monitoring tools such as Nagios and Munin, or those provided by the (operating) system were mentioned 
(73\,\%). 
As opposed to this, only 
15\,\% of the
participants reported that they are using a (commercial) APM tool. Three participants reported about self-developed tools, which seems to be a current trend to use general-purpose data analytics and visualization components (e.g., logging and Graphite) to set up custom monitoring infrastructures. 



\subsection{Few practitioners use performance models, despite widespread interest}
\label{sec:no_perf_model}

The results of our survey reveal that performance models are currently not used in industry and there appears to be no trend towards their adoption either (see Fig.~\ref{fig:perf_models}). Our survey shows that 
88\,\% of the participants do not apply models for performance management, even though 18 (almost 70\,\%) of them state that they would like to use such models. While most participants are aware of performance modeling formalisms, their knowledge seems to be shallow, since our results show that only 5 (19\,\%) of the participants have (some) knowledge about queuing networks, i.e., the most well-known performance modeling formalism.

\section{Implications of our findings}
\label{sec:implications}
As discussed in Section~\ref{sec:regularly}, most surveyed companies do not regularly conduct performance evaluations. In prior work, Leitner and Bezemer~\cite{Leitner2017} showed that in most open source projects performance evaluations are not conducted on a regular basis either. These findings suggest that there is a mismatch between what the plethora of performance engineering research delivers, and what practitioners are really looking for. Below, we discuss the most important implications of our study for researchers.

\subsection{The complexity of performance engineering approaches is a barrier for wide-spread adoption by practitioners}
\label{sec:impl_complexity}

Software performance assurance activities are complex tasks by nature that require much knowledge of various aspects of the entire software life-cycle. As a result, performance engineering approaches, which are often highly complex, are not straightforward for practitioners to adopt and understand. For example, performance modeling is a widely leveraged technique in research that can be particularly suitable in a DevOps context. As performance tests can be conducted much faster on performance models than on real applications, performance models could work well for applications that release many times per day. Unfortunately, Section~\ref{sec:no_perf_model} shows that the application of performance models is rare in industry. The lack of participants’ knowledge is the most likely cause for not having a clear opinion about the underlining science of such models. Performance modeling techniques, being mostly research prototypes, often lack documentation and require expert knowledge to be leveraged, which makes their integration for non-experts tedious. Hence, the valuable outcomes of the performance models may be difficult for practitioners to interpret, digest, or even trust. 


\subsection{Performance engineering approaches must be lightweight}
\label{sec:impl_lightweight}

Our findings highlight the need for more lightweight performance engineering approaches, which still retain the necessary accuracy. A step towards such approaches might be automating aspects of existing approaches and hiding their associated complexity from the practitioner. The high amount of required effort upfront to construct and calibrate a performance engineering technique (e.g., performance modeling) may be an extra barrier for industrial adoption. While academic studies show the benefits of performance models for reasoning about design decisions and trade-offs~\cite{reussner2016modeling}, industry may fear the high upfront cost. 

In addition, automated and systematic performance engineering approaches, e.g., creating and updating performance models,  may facilitate the adoption of such techniques in industry. While automated extraction approaches approaches already exist~\cite{SPEC-RG-2015-001-DevOpsPerformanceResearchAgenda}, there is still no ``one-click'' solution, which would significantly reduce the entry barrier. One important step to enable more lightweight performance engineering approaches. An example of tools that aim at reducing the entry barrier are APM tools. Unfortunately we did not observe wide-spread adoption of such tools by practitioners, yet. 

\subsection{Performance engineering approaches must smoothly integrate with existing tools in the DevOps pipeline}
\label{sec:impl_integrate}
A possible explanation for the low adoption of performance engineering practices in DevOps
could be that performance engineering approaches are typically not designed with the consideration of DevOps as a general context.
On the other hand, existing tools that are used in many DevOps settings, such as Puppet and Docker, do not integrate nicely with existing performance engineering processes in industry. For example, many practitioners still rely on old-fashioned tools, such as SSH and system tools, to deploy their applications and monitor performance. In addition, we observed that even though many participants conduct application level monitoring, they do so without the use of specialized tools (such as APM tools). 

Our recommendation for performance engineering researchers is to ensure that their tools integrate smoothly in existing DevOps pipelines. For example, we observed in the survey responses that Jenkins CI is by far the most popular CI tool. Hence, we recommend that researchers provide plugins that allow an easy integration of their performance evaluation tools in Jenkins.

\section{Threats to Validity} 
\label{sec:threats}
In this section, we discuss the threats to validity of our study.

\textbf{Internal validity.} Threats to internal validity relate to the participant bias and errors. A first internal validity threat concerns the possible selection bias for survey participants. To avoid such bias, we advertised the survey in a wide variety of channels (see Section~\ref{sec:advertisement}). However, some of these channels (e.g., the SPEC mailing list) may target a specific audience. Hence, the results of our survey may be biased. In addition, our survey targeted industrial projects, which are mostly closed-source. Hence, our findings do not necessarily extend to open source projects. Future studies are necessary to further explore how performance is addressed in DevOps in other companies and in open source projects.

\textbf{Construct validity.} A threat to the construct validity of this study is that our survey consisted mostly of closed-ended questions. As a result, the richness of the responses may be affected. However, we felt that the advantages of closed-ended questions outweighed the disadvantages: closed-ended questions are easier to answer and analyze~\cite{burns2008basic,fink2012conduct,salant1994conduct,openquestions2003}. Hence, we focused on closed-ended questions.

\section{Conclusion} 
\label{sec:concl}

In this paper, we highlight the results of an independent survey that focused on performance engineering practices in DevOps. 
We found that two third of participants do not conduct performance evaluation on a regular basis, and among the ones that conduct performance evaluations, 50\,\% of the participants spend less than 5\,\% of their time on them. 
For what concerns the applied practices in DevOps, most participants perform continuous integration, while continuous deployment and continuous provisioning is seldom implemented. 
Tool-wise, Jenkins is the most used end-to-end tool for implementing DevOps practices. 
We also found that the use of performance models by practitioners is very low.


One explanation for the low adoption of performance engineering practices in DevOps could be that the DevOps movement is still in its infancy, and developers are still getting used to the opportunities that this movement offers in terms of automation of performance engineering processes.

Our survey shows that even though the adoption of DevOps is relatively widespread in industry, performance engineering practices are lagging behind. Future research should focus on assisting software developers and performance engineers to convert their existing performance engineering practices into the DevOps pipeline.

\section*{Acknowledgements}
This research was conducted by the SPEC RG DevOps Performance Working Group.\footnote{https://research.spec.org/devopswg} We would like to thank all survey participants for their responses.
The authors have benefited from discussions with various colleagues during community events such as the Dagstuhl seminar on ``Software Performance Engineering in the DevOps World''~\cite{vanHoornJamshidiLeitnerWeber2017ReportFromGIDagstuhlSeminar16394}.

\medskip

\bibliographystyle{IEEEtranS}


\end{document}